\documentclass[12pt,preprint]{aastex}

\shorttitle{Stellar Parameters of GGSS Stars}
\shortauthors{Bizyaev et al.}

\begin{document}

\title{
The Space Interferometry Mission Astrometric Grid Giant-Star Survey. III.
Basic Stellar Parameters for an Extended Sample.
}

\author{Dmitry Bizyaev\altaffilmark{1,2,3}, Verne V. Smith\altaffilmark{4},
Katia Cunha\altaffilmark{4,5,6}}
\email{dmbiz@apo.nmsu.edu, vsmith@noao.edu, cunha@noao.edu}

\altaffiltext{1}{New Mexico State University and Apache Point Observatory,
Sunspot, NM, 88349}
\altaffiltext{2}{Sternberg Astronomical Institute, Moscow, 119899, Russia}
\altaffiltext{3}{Isaac Newton Institute of Chile, Moscow Branch}
\altaffiltext{4}{National Optical Astronomy Observatory, Tucson, AZ, 85719}
\altaffiltext{5}{Steward Observatory, University of Arizona, Tucson, AZ  85719} 
\altaffiltext{6}{on leave, Observatorio Nacional, Rio de Janeiro, Brazil}

\begin{abstract}
We present results of high resolution ($\sim$ 55000) spectral observations
of 830 photometrically pre-selected candidate red giants in the magnitude range
of  V = 9-12.
We develop a pipeline for automated determination 
of the stellar atmospheric parameters from these spectra and
estimate  $T_{eff}$, $\log~g$, [Fe/H], microturbulence velocity, and projected
rotational velocities, $vsini$, for the stars.  The analysis confirms that the candidate
selection procedure yielded red giants with very high success rate.
We show that most of these stars are G and K giants with slightly subsolar
metallicity ([Fe/H] $\sim$ -0.3 dex).
An analysis of Mg abundances in the sample results in consistency of the 
[Mg/Fe] vs [Fe/H] trend with published results. 
\end{abstract}


\section{Introduction}

The photometric Grid Giant Star Survey \citep[GGSS;][]{patterson01, M00}
was designed to search for G and K giant stars in selected relatively large
areas ($\sim$ 1 sq. degree), more or less evenly spaced all around the sky. 
A number of GGSS giants were selected as potential candidates to be
astrometric reference grid stars for the NASA Space Interferometry 
Mission (SIM).
The primary motivation for the spectroscopic follow-up survey of GGSS
candidate giants was to monitor the line-of-sight velocity at a high
level of precision, using an iodine cell to identify 
radial velocity - stable stars in support of SIM.  
During the radial velocity monitoring campaign of the
Northern hemisphere sample at the McDonald Observatory (University of Texas), we
obtained some 3000 spectra for more than 1500 candidates with 
magnitudes V $\sim$ 9 -- 12. 
Although an iodine cell was used to imprint
an I$_{\rm 2}$ spectrum onto the stellar spectrum in order to monitor 
radial velocities, 
a significant number of spectra were obtained without the cell; these  
have moderate signal-to-noise (S/N), and are suitable for an abundance analysis. 
The spectroscopic analysis of this large number of spectra is the object of 
this paper
which discusses an automated analysis for deriving stellar parameters 
and metallicities.

We have previously analyzed, using an automated technique, the stellar 
parameters 
for an earlier sample of SIM reference grid
star candidates \citep{b06}.  The stars considered in the present 
paper have somewhat brighter apparent magnitudes, because the selection criteria 
were changed to accommodate revised SIM technical requirements, and in general
have higher S/N (typically S/N $\sim$ 30 - 50).
Also, in contrast to our earlier study \citep{b06}, we employ here a different and improved
technique to derive the parameters using an approach similar to that developed by \citep{TGMET},
which is found to be better applicable to spectra with S/N of moderate quality (of the order of tens). 
In this paper, we focus on the characterization of the GGSS sample with the construction of
a pipeline for the derivation of the target stars atmospheric parameters and metallicities (iron 
abundances). Such results will add to the large database of metallicities
which have been produced over recent years by large surveys using low-resolution spectra (e.g. SDSS) 
\citep{lee081,lee082,allende08}. The high-resolution spectra in this study, 
in contrast with surveys which analyze lower resolution spectra, 
will allow future studies to determine the abundances of several chemical elements with accessible 
spectral lines in the wavelength range covered by these spectra.

\section{Observations}
\label{spectroscopy}

For the preparatory RV campaign for SIM, we conducted 
spectroscopic observations of a large sample (1500+ objects) of 
photometrically pre-selected
candidate red giants. The whole-sky sample is a mix of the brightest GGSS
stars \citep{frink01,patterson01}, as well as Tycho stars
\citep{catanzarite04,ciardi04,thompson05}. Note that this joint sample
has only a very small overlap with that studied by Bizyaev et al. (2006) because of the 
revised SIM technical requirement to use brighter reference stars.

Observations were conducted using the 2.1m telescope at the McDonald 
Observatory and the
Sandiford Cassegrain echelle spectrograph, which provides a nominal resolution 
R = 55000 (McCarthy et al. 1993).
In contrast to the preceding radial velocity monitoring
campaign \citep{b06}, the observations discussed here employed the iodine cell 
technique \citep[e.g.][]{butler96}. The gas 
cell \citep[provided by W. Cochran, ][]{cochran94}
with evaporated iodine was heated to +50\degr C and 
placed in the beam before the slit of
the spectrograph. For most of the stars, observations were made with and
without the iodine cell. The latter frames were used as
templates for the radial velocity measurements. 
The radial velocity monitoring results will be the subject of a future paper. 
Here we focus on the echelle spectra obtained without the iodine cell and which
will be used for abundance analyses.  We observed 841 spectra for 830 program stars. 
In addition, we obtained spectra of 10 very bright stars with previously
identified atmospheric parameters for comparison purposes. 
A day-light solar spectrum is also included in the analysis.

Th-Ar comparison spectra were taken right before each program star.
The adopted instrumental setup enabled us to cover the spectral range from 5000-5900 \AA\, .
There were 24 spectral orders across the chip, each order being 1200 pixels long.
The first and last (as a rule incomplete) orders were rejected, as well as 
three orders in the middle of the CCD chip that are contaminated by an internal reflection. 
The mean S/N for all spectra was 42, with an r.m.s. of 12, and the typical
exposure time was 1000 sec. Observations of the bright comparison stars resulted in spectra
with S/N of 150 or more.

The observed spectra were reduced from two-dimensional to ``echelle" format
in a standard way with the IRAF\footnote{IRAF is distributed by 
NOAO which is  operated by AURA, Inc. under contract with the NSF.} 
software package. Corrections for
bias, flat field, and scattered light were applied, and cosmic ray hits 
cleaned out by the IRAF's tasks from {\it crutil} package. 
The internal accuracy of the wavelength calibration via the Th-Ar reference
spectra was on average 0.001 \AA.

Intermediate resampling introduces systematic errors in the wavelength 
calibration and degrades the RV accuracy.
We preserve the initial sampling and
do not concatenate the echelle orders into a single 1-D spectrum. 
Rather than that, we keep each echelle order separate, with
its individual sampling.

\section{The TAGRA Pipeline}

We designed an automated pipeline to derive the effective Temperatures, iron Abundances, and
surface GRAvities (TAGRA) for the sample stars via comparisons of observed spectra with synthetic ones. 
Near-IR photometric data were also used in the pipeline (see below). 

\begin{figure}
\includegraphics[scale=.80]{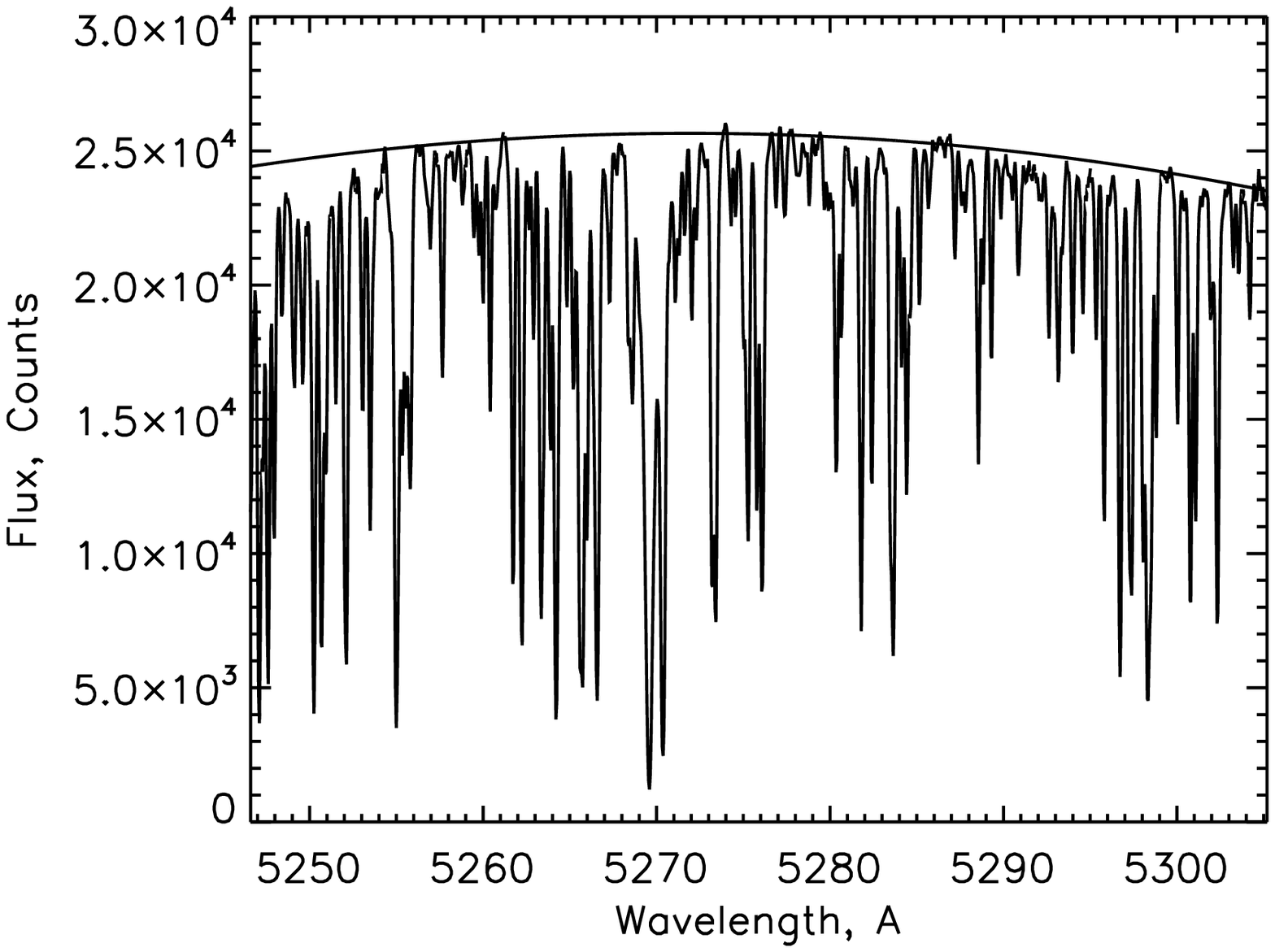}
%--%\epsscale{.70} 
%\plotone{fig1.eps}
\caption{An echelle order of the Arcturus spectrum and the 
polynomial fitting defining the continuum (solid line).
\label{f_cont}
}
\end{figure}

Observed spectra were normalized to an empirically defined continuum level. 
We used the ultra-high resolution spectra of
both the Sun and Arcturus from the atlas by \citet{yellowbible}, 
degraded to our resolution, in order to identify pieces of spectra relatively free of 
spectral lines and which could be used as continuum points. For each spectral order we
used as many continuum points as possible; Table~1 lists 235 such continuum points
which were included in the pipeline.  
The continuum points in our observed spectra were fit with low-order polynomial functions. 
Figure~\ref{f_cont} shows an example of the continuum fitting within
one spectral order of the degraded Arcturus spectrum, whose spectrum
is quite representative of the stars analyzed here.

\begin{deluxetable}{ llll }
\tablewidth{0pt}  
\tablecaption{    
Normalized pseudo-continuum level between 5000 and 6000\AA.
}
\tablehead{
$\lambda(start)$ & $\lambda(end)$ & Mean Cont.(Arcturus and Sun) & r.m.s. \\       
} 
\startdata
4993.10 & 4993.20 & 0.99240 & 0.00099\\
5003.50 & 5003.60 & 0.98205 & 0.01336\\
5009.00 & 5009.10 & 0.97865 & 0.00771\\
5015.60 & 5015.75 & 0.97645 & 0.00474\\
5018.90 & 5019.00 & 0.98005 & 0.01690\\
\enddata
\tablecomments{The wavelength range (in \AA) and the 
normalized continuum level in the spectra of Arcturus and the Sun. 
Table~1 is published in its entirety in the electronic
edition.
}
\label{tab1}
\end{deluxetable}

\subsection{Effective Temperatures}
\label{Teff}

The introduction of one more degree of freedom
($T_{eff}$ in this case) into the modeling and fitting using synthetic spectra
leads to larger uncertainties in the derived stellar parameters.
Since all program stars were observed by 2MASS 
\citep{2mass}, this provides additional photometric
information which can be used to constrain effective temperatures for the pipeline.
Similar to the analysis by  \citet{b06}, effective temperatures were
determined from the 2MASS colors (J-K). 
We use a calibration by \citet{alonso99} and transformation
equations from 2MASS to Alonso's TCS color system given by
\citet{ramirez05}. The galactic foreground extinction is taken from the 
GGSS and Tycho sample papers \citep{patterson01,ciardi04}. We note that this
calibrations of (J-K) for giants does not have a dependence on the metallicity.
Since the program stars are relatively bright, this calibration provides
good accuracy for the effective temperature, typically better 
than 125 K (Alonso et al. 1999). 

\subsection{Iron Lines, Surface Gravities and Microturbulent Velocities}
\label{logg}

With effective temperatures defined by the 2MASS colors, iron lines (Fe I and Fe II)
were used to determine the remaining stellar parameters: surface gravity and
microturbulent velocities.  Potential Fe lines were taken from the studies by
% was fixed, we select relatively narrow 
%(of the order of 1\AA) ranges of spectra
%which contain iron lines suitable for the estimation of [Fe/H] and
%relatively free of other lines. We start with lists of lines from
\citet{fulbright06} and \citet{b06}.  The spectral regions containing these Fe I and
Fe II lines were investigating for their suitability as Fe abundance indicators by
using the complete linelists (containing all lines from all species from
Kurucz \& Bell 1995)
and generating synthetic spectra with 
the MOOG spectrum synthesis code \citep{MOOG} and ODFNEW Kurucz's
atmospheric models \citep{kurucz}.  A small spectral window around each Fe line
was studied as to how the line responded to 
variations in the surface gravity, iron abundance, and microturbulence 
($\log~g$, [Fe/H], and $\xi$). 
Spectra in each of the selected regions were inspected and any heavily blended 
Fe lines were excluded. Also very weak lines that were not well-defined 
in our medium quality spectra were excluded from the final analysis.
We ended up with 81 spectral regions; the list of wavelength ranges and the Fe 
lines in each of them is presented in Table~2.  
The excitation potentials and log(gf) values for the sample Fe lines are adopted from 
\citet{b06} and \cite{fulbright06}.

\begin{deluxetable}{ llll }
\tablewidth{0pt}  
\tablecaption{
The spectral  window around the iron lines utilized in TAGRA.
}
\tablehead{
$\lambda(start)$ & $\lambda(end)$ & line &  $\lambda$, \AA \\
} 
\startdata
5054.55 & 5054.90 &  FeI & 5054.643 \\
5141.45 & 5142.05 &  FeI & 5141.739 \\
5236.00 & 5236.30 &  FeI & 5236.202 \\
5307.00 & 5307.70 &  FeI & 5307.361 \\
5412.65 & 5412.90 &  FeI & 5412.784 \\
\enddata
\tablecomments{
Table~2 is published in its entirety in the electronic edition.
}
\label{tab2}
\end{deluxetable}

%Using the effective temperature from \S~{\ref{Teff}} and typical expected 
%$\log~g$, [Fe/H], $\xi$, and $vsini$ as 2.0, -0.1, 2.0, and 2.0,
%respectively, we generate a
Radial velocities are determined as part of the spectral analysis and
initial measurements of the stellar radial velocities were determined
from synthetic spectra calculated for the region around the
Mg I 5180\AA~ line. The synthetic spectra were then convolved with a Gaussian
instrumental profile, which corresponded to the typical FWHM of the
observations (2.4 pixels); the radial velocity (RV) of the star was then determined
via cross correlation of synthetic and observed spectra. This RV was used as the
initial radial velocity in further modeling. 
%It is different from the resulting RV by 1.25 km/s, on average.

As the next step in the analysis of a given star, synthetic spectra were computed
for each spectral window in Table~2 and were fit to the observed 
spectra using as free parameters RV, $\log~g$, [Fe/H], $\xi$, and $vsini$. 
%Note that a few lines fall into the ranges from Table~2 more than once 
Note that since the echelle orders have overlapping wavelengths, a few 
sample lines 
were observed in more than one order.
To avoid using low S/N regions that can occur near the edges of echelle orders, 
we rejected fitting a line if it was closer than 100 pixels to the edge.
Variations in the width of the instrumental profile (IP hereafter) were also 
investigated as an additional model parameter.
Because the spectrograph focus was monitored regularly during  the observations 
with Th-Ar line profiles, we found that the IP was stable enough 
during observations, so that it could be kept fixed for each fit.
We tried the Powell, Levenberg-Marquardt, as well as the downhill-simplex 
(also called Amoeba) optimization algorithms \citep{numrecC} 
to derive the best-fit parameters. The optimized values are the chi-square
derived from the synthetic and observed spectra for each considered line
region. Tests indicated that, although being the slowest of the three algorithms
tested, the downhill-simplex algorithm yielded realistic best-fit parameters, 
whereas the Powell and Levenberg-Marquardt algorithms often failed to find the minimum.

We considered two approaches for this analysis: 1) doing the interpolation between the model atmospheres 
to generate spectra for each set of parameters versus 2) doing the interpolation
between synthetic spectra generated with a fixed grid of parameters. The latter 
approach is preferable for our computations as it is much faster 
and requires less computer resourses since we use a limited number of short 
pieces of spectra.
% whereas the former approach seems to be more correct methodologically.
In both cases we synthesized pieces of spectra in the selected wavelength region.
The synthetic spectra were computed using
the LTE spectrum synthesis code MOOG \citep{MOOG}.
% \citep{kurucz}.

For the first method, synthetic spectra were generated via an interpolation  
between the Kurucz atmospheric models \citep{kurucz} with parameters varied
until a best fit between synthetic and observed spectra was obtained.
In the second approach, we produced an entire grid of synthetic spectra
in the selected wavelength ranges
for $T_eff$ =3500 -- 6000 K, $\log~g$ = 0 -- 5 dex, [Fe/H] = -2.5 -- +1.0 dex, 
and $\xi$ = 0.5 -- 3.5 km/s with steps of
250 K, 0.5 dex, 0.5 dex, and 1.0 km/s, respectively for Teff, log g, metallicity 
and microturbulent velocity. 
Real spectra were then interpolated between the synthetic spectra and these results
were compared to those obtained directly from synthesis. If a linear
interpolation was applied, the difference between the spectra generated using
the two approaches could be as close as 1\%.   Using a second-order polynomial
interpolation, the difference becomes less than 0.5\%, and thus we expect to see no
significant difference if spectra from both methods are compared with real
spectra of intermediate to good quality (S/N up to 200). 
We conclude that use of the second-order polynomial interpolation yields
accuracies sufficient for values of the S/N up to about 200.   We used the
second-order polynomial interpolation between the pre-compiled spectra in
the final pipeline.  

Projected rotational velocities, $vsini$, were measured from fits to 
all spectral lines falling within the echelle order near $\lambda5740\AA$ 
(with rejection of 100 pixels from the edges of the order).  
A synthetic spectrum was generated using the best-fit model parameters 
for each star and this
spectrum is broadened over a range of values of $vsini$), with the best fit 
used to define $vsini$.  We note that at the spectral resolution of
R=55000, this broadening which is fit as rotational will include 
underlying macroturbulence within the star as well, thus any values
of $vsini \, \le $8 km-s$^{-1}$ should be considered as upper limits.
At the same time, the rotational velocities for the Sun and Arcturus
derived by TAGRA are very close to their real values. Our attempts to fit 
the spectra with the rotational and macroturbulence
broadening simultaneously, both directly and in Fourier space
\citep[see][]{gray76}, failed because of insufficient S/N and a lack of
clearly isolated lines in the spectra of our stars. An alternative approach
to the $vsini$ estimates using the cross-correlation technique for our 
stars is developed by \citet{carlberg10} and results in the 
discovery of a few rapidly rotating stars in our sample. 
We postpone a more detailed discussion of the rotational and 
mactoturbulence broadening until the next paper in this series.

\begin{figure}
\includegraphics[scale=.80]{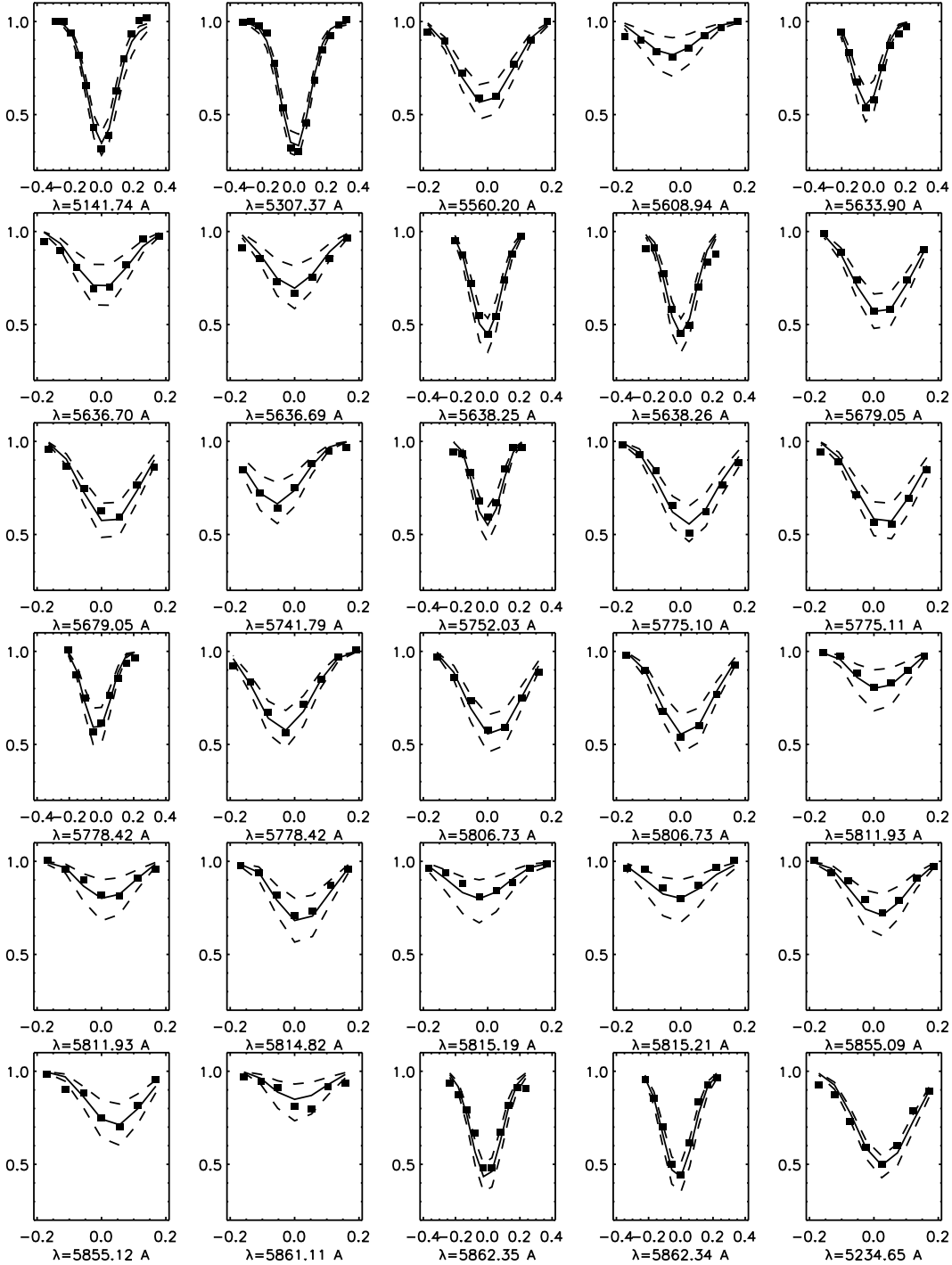}
\caption{
Observed and best-fitting synthetic profiles for a sample of Fe lines  in Tyc2630-01070-1,
a typical star from our sample. The squares indicate the observed spectra, the
solid curve shows the best-fit synthetic model, and the dashed curves are
offset by $\pm$ 0.5 dex in [Fe/H].
\label{f_fit}
}
\end{figure}

The final atmospheric parameters for the sample stars 
as well as the estimated $vsini$
are given in Table~3.
The resulting best fiting spectra to narrow spectral regions around 
selected iron lines for a typical star in our sample are shown in 
Figure~\ref{f_fit}. The squares indicate the observing spectra, the
solid curves show the best-fitting synthetic models, and the dashed curves
represent [Fe/H] offsets by $\pm$ 0.5 dex.

\begin{deluxetable}{ lcccccccccc }
\tablewidth{0pt}  
\tablecaption{The atmospheric parameters derived with TAGRA}
\tablehead{  
Name & RA(J2000) & DEC(J2000) & $m_B$ & $A_V$ & $T_{eff}$ & $\log~g$ &
$[Fe/H]$ & $\xi$ & $vsini$ & S/N\\
 & & & mag & mag & K & dex & dex & km/s & km/s &  \\
} 
\startdata
G0000+67.16328 & 14:59:18.5&  67:34:02 & 10.80 & 0.49 & 4624 & 2.32 & -0.53 & 1.54 &  2.46 &   37\\
G0024+61.2521  & 00:15:06.6&  62:26:51 & 10.58 & 0.45 & 4489 & 1.97 & -0.36 & 1.41 &  2.37 &   38\\
G0024+61.6685  & 00:15:22.2&  62:18:15 & 10.76 & 0.47 & 4677 & 3.84 &  0.08 & 0.50 &  0.62 &   28\\
G0024+61.9144  & 00:18:18.2&  62:13:14 & 10.42 & 0.43 & 4797 & 2.34 & -0.11 & 1.49 &  1.85 &   34\\
G0112+61.6061  & 00:47:22.8&  62:16:03 & 10.69 & 0.47 & 4645 & 3.36 & -0.54 & 0.94 &  5.68 &   14\\
G0121-16.44    & 00:52:18.1& -16:31:03 & 10.46 & 0.06 & 4684 & 2.92 & -0.21 & 1.30 &  1.26 &   34\\
\enddata
\tablecomments{Name, equatorial coordinates for epoch (J2000.0),
B-magnitude $m_B$, extinction $A_V$, and derived atmospheric parameters are shown. 
Table~3 is published in its entirety in the electronic edition.
}
\label{tab3}
\end{deluxetable}

%\subsection{Stellar Rotation}}

\subsection{Reliability of Derived Stellar Parameters}

The external accuracy of the derived stellar parameters
in this study can be evaluated from independent determinations 
of stellar parameters (all using TAGRA) for a sub-sample of 11 
stars which have been observed two times during the Survey. 
Table~4 summarizes the mean values of $\log~g$, [Fe/H], $\xi$, and
$vsini$, and standard deviations, obtained for these stars. It can be seen that the
typical accuracy is 0.24 dex and 0.05 dex for $\log~g$ and [Fe/H], and
0.09 and 0.17 km/s for $\xi$ and $vsini$.
The internal precision of the derived parameters can be estimated
by evaluating each parameter in all selected spectral regions by
varying them one at a time.  We found that the formal internal
accuracy is less than external accuracy.
%Table~4. Uncertainties of the Atmospheric Parameters Estimated From Multiple
%Observations.

\begin{deluxetable}{ lrrrr }
\tablewidth{0pt}  
\tablecaption{
Mean stellar Parameters and uncertainties obtained from two-epoch observations.
}
\tablehead{
Name & $\log~g$ & $[Fe/H]$ & $\xi$ & $vsini$ \\
}
\startdata
Tyc2773-01666-1 & 2.930 $\pm$ 0.238 & -0.279 $\pm$ 0.021 & 1.583 $\pm$ 0.091 & 1.68 $\pm$  0.16\\
Tyc2269-01094-1 & 3.115 $\pm$ 0.064 & -0.405 $\pm$ 0.031 & 0.549 $\pm$ 0.049 & 3.71 $\pm$  0.48\\
Tyc2185-00133-1 & 2.774 $\pm$ 0.340 & -0.657 $\pm$ 0.016 & 1.331 $\pm$ 0.063 & 5.93 $\pm$  0.01\\
Tyc1726-00742-1 & 2.347 $\pm$ 0.177 & -0.294 $\pm$ 0.042 & 1.333 $\pm$ 0.001 & 1.94 $\pm$  0.20\\
Tyc1780-00654-1 & 2.444 $\pm$ 0.354 & -0.436 $\pm$ 0.111 & 0.953 $\pm$ 0.248 & 4.63 $\pm$  0.07\\
Tyc0001-00818-1 & 2.943 $\pm$ 0.257 &  0.161 $\pm$ 0.060 & 1.228 $\pm$ 0.210 & 0.00 $\pm$  0.00\\
G0410+11.148    & 1.785 $\pm$ 0.327 & -0.324 $\pm$ 0.077 & 1.250 $\pm$ 0.057 & 3.12 $\pm$  0.29\\
G1705+22.14898  & 2.267 $\pm$ 0.299 & -1.212 $\pm$ 0.049 & 1.895 $\pm$ 0.048 & 5.07 $\pm$  0.36\\
G2238+33.1387   & 2.940 $\pm$ 0.058 &  0.132 $\pm$ 0.044 & 1.031 $\pm$ 0.088 & 0.10 $\pm$  0.10\\
Tyc1753-00655-1 & 1.752 $\pm$ 0.363 & -0.319 $\pm$ 0.067 & 1.385 $\pm$ 0.034 & 1.50 $\pm$  0.18\\
Tyc3973-02373-1 & 2.151 $\pm$ 0.145 & -0.232 $\pm$ 0.056 & 1.411 $\pm$ 0.093 & 1.39 $\pm$  0.04\\
Mean            &             0.238 &              0.050 &             0.089 &             0.17 \\
\enddata
%\tablecomments{The mean and r.m.s. of the atmospheric parameters 
%from multiple independent measurements of 11 program stars.
%}
\label{tab4}
\end{deluxetable}

%% 0.238364    0.0521818    0.0892727     0.172000

\begin{figure}
\includegraphics[scale=.80]{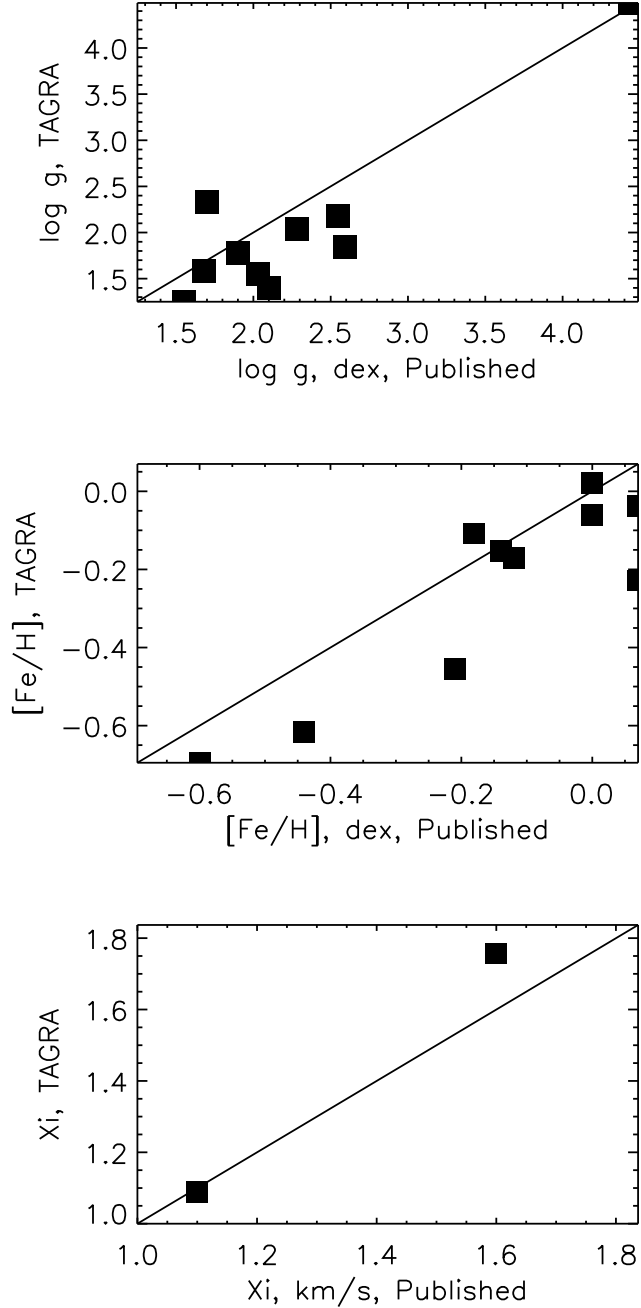}
%--%\epsscale{.70} 
%\plotone{fig1.eps}
\caption{Comparison between the derived surface gravities ($\log~g$), metallicities
([Fe/H]) and microturbulent velocities ($\xi$) with published values from the literature 
for 10 bright stars which were observed as a control sample. The literature values are
from  Souburan et al. (2008), Luck \& Challener (1995), McWilliam (1990) and Brown et al. (1989).
\label{bright_stars}
}
\end{figure}

In addition, high-quality spectra for a sample of bright and well studied stars
were obtained under the same conditions and with the same instrument and setup as 
all other program stars studied here; these spectra had signal-to-noise ratios 
better than $\sim$ 150. The stellar parameters derived here with TAGRA 
(and listed in Table \ref{comp_bs})
can be compared with results from classical spectroscopic analyses previously 
published in the literature 
(listed in Table \ref{comp_bs}): Arcturus \citep{soubiran08}, $\beta$Oph \citep{luck95},
$\gamma$Dra, $\upsilon$Dra, $\pi$Her, 11 Umi, HD160290, HD218029 
\citep[all from ][]{mcwilliam90}, HD152812 \citep{brown89}, as well as the Sun. 
Figure~\ref{bright_stars} shows the comparison between the published stellar
parameters and those derived with TAGRA.
The mean difference in $\log~g$, [Fe/H], and $\xi$
within the whole sample is -0.24 $\pm$ 0.40 dex, -0.10 $\pm$ 0.12 dex, 
and 0.07 $\pm$ 0.12 dex, respectively. Note that for 
the Sun and Arcturus in particular, the differences  (TAGRA - published) for
the $\log~g$ and [Fe/H] is only -0.04 and -0.08 dex, respectively.
These results are also shown in Table ~\ref{comp_bs}.

\begin{deluxetable}{ lrrrrrrr }
\tablewidth{0pt}  
\tablecaption{
Comparison of derived parameters for published values of iron abundance [Fe/H], surface gravity
$\log~g$, and microturbulence $\xi$ to those obtained with TAGRA in this
paper.
}
\tablehead{
Star & $T_{eff}$ publ.& [Fe/H] publ. & $\log~g$ publ. & $\xi$ publ &
[Fe/H] & $\log~g$ & $\xi$ \\
 & K & dex & dex & km/s & dex & dex & km/s \\
} 
\startdata
Sun       & 5778 &  0.00 & 4.44 & 1.10 & -0.06 & 4.49 & 1.09 \\
Arcturus  & 4280 & -0.60 & 1.90 & 1.60 & -0.69 & 1.78 & 1.76 \\
$\gamma$ Dra   & 3830 & -0.14 & 1.55 &  --  & -0.15 & 1.25 & 1.62 \\
11 Umi         & 4120 &  0.07 & 2.03 &  --  & -0.23 & 1.55 & 1.69 \\
$\beta$ Oph    & 4475 &  0.00 & 1.70 &  --  &  0.02 & 2.33 & 1.53 \\
HD152812  & 4150 & -0.44 & 2.10 &  --  & -0.62 & 1.40 & 1.56 \\
HD160290  & 4440 & -0.21 & 2.59 &  --  & -0.46 & 1.84 & 1.44 \\
HD218029  & 4280 &  0.07 & 2.28 &  --  & -0.04 & 2.04 & 1.58 \\
$\pi$ Her      & 4100 & -0.18 & 1.68 &  --  & -0.11 & 1.59 & 1.84 \\
$\upsilon$ Dra & 4520 & -0.12 & 2.55 &  --  & -0.17 & 2.18 & 1.53 \\
\enddata
\tablecomments{
Name of the star, its effetive temperature, and published 
[Fe/H], $\log~g$, and $\xi$ versus those obtained with TAGRA.
}
\label{comp_bs}
\end{deluxetable}

As a further test, we also performed a "classical" spectroscopic analysis 
for one of the bright stars, $\beta$Oph, using the same list of lines 
as in the pipeline analysis. If a similar approach is adopted of
using the effective temperature from the photometric calibration as a fixed
value,  the other derived parameters are: $\log~g$ = 1.3 dex, [Fe/H] = +0.02
dex, and $\xi$ = 1.9 km-s$^{-1}$, i.e. in good agreement with those found with
the TAGRA. 
%% No !!!
%%Note that the metallicity derived with the photometric $T_{eff}$ in the
%%classical analysis is very close to that estimated with the TAGRA.
Note that these parameters were determined under an assumption of 
published $T_{eff}$.
If the effective temperature is also derived
spectroscopically (from the excitation equilibrium of Fe I lines)
we obtain: Teff=4630 K, $\log~g$ = 1.9 dex, [Fe/H] = +0.15 dex, and $\xi$ = 1.9 km/s.

% All together
% Log g  :    -0.237200 +-      0.396377
% [Fe/H] :   -0.0955000 +-      0.116410
% ksi, exist:    0.0735000 +-      0.118087
%
%Sun and Arcturus only:
%Log g  :   -0.0350001 +-      0.117380
%[Fe/H] :   -0.0775000 +-     0.0247487
%ksi    :    0.0735000 +-      0.118087

\begin{figure}
\includegraphics[scale=.80]{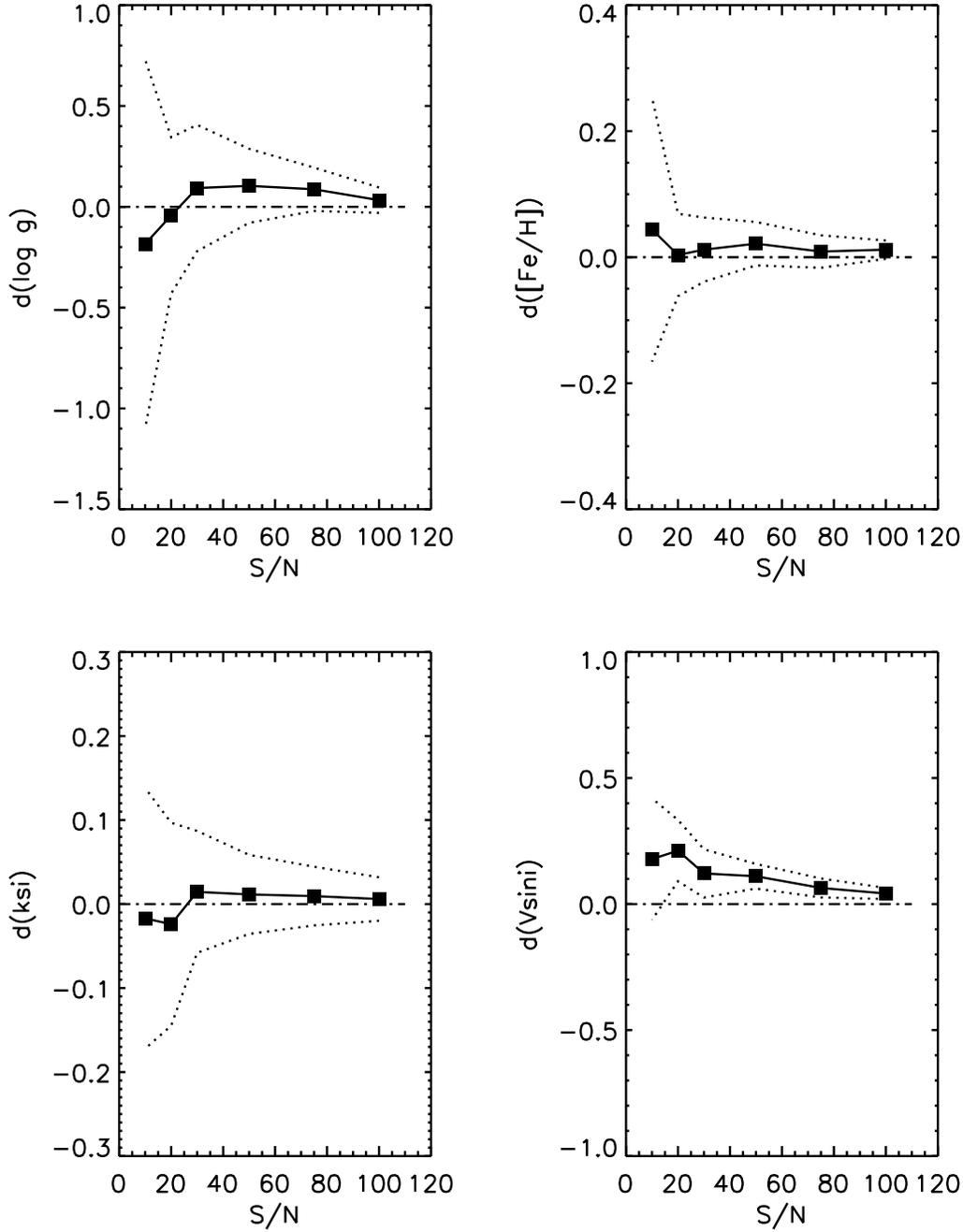}
%--%\epsscale{.70}
%\plotone{fig2.eps}
\caption{
Differences between the parameters $\log~g$, [Fe/H], $\xi$, and $vsini$ 
estimated from the noise-added spectra of Arcturus and those from high-S/N 
spectra (see explanations in text). The solid line with squares shows the
difference. Zero at the vertical axes on all panels in
Figure~\ref{sp_artificial} is designated by the dash-dotted horizontal line.
Dotted lines designate the r.m.s. scatter of
the parameters determined from the deteriorated spectra.
\label{sp_artificial}
}
\end{figure}

Finally, the accuracy of the TAGRA pipeline was tested from an analysis
of the spectrum of Arcturus degraded with the addition of artificial noise. 
We performed Monte Carlo simulations and derived the basic stellar parameters 
for spectra with a range in S/N from 10 to 100, and produced sets of thirty 
noisy artificial spectra for various values of S/N. 
Figure~\ref{sp_artificial} shows the difference between the parameters
$\log~g$, [Fe/H], $\xi$, and $vsini$ estimated from the original high-S/N 
spectra and the same spectra with added noise. 
The solid line with squares shows the
difference (zero at the vertical axes in all panels in
Figure~\ref{sp_artificial} is designated by the dash-dotted line and 
corresponds to the case of no difference). 
Dotted lines designate the r.m.s. scatter of
the parameters determined from the degraded spectra. 
As can be seen from Figure~\ref{sp_artificial}, the spectrum quality 
imparts rather small uncertainties in the metallicity,
microturbulence and $vsini$ (0.05 dex, 0.05 dex, and 0.2 km/s,
respectively), with somewhat larger uncertainties in the surface gravity (0.25 dex).
The $vsini$ and log g also exhibit some moderate systematic differences 
for the degraded spectra (of the order of 0.1 km~s$^{-1}$ and +0.1 dex, respectively). 

Considering the reliability of basic atmospheric parameters derived using 
the TAGRA pipeline, we conclude that the metallicity ([Fe/H]), microturbulence
($\xi$), and $vsini$ are the best-defined parameters. 
The surface gravities have poorer accuracies and may be affected somewhat by systematic 
errors. This is not unexpected because the primary source of $\log~g$ is the balance 
between the FeI and FeII lines. There are not as many FeII lines as FeI 
in this wavelength range and, in addition, the available FeII lines are 
concentrated towards the more line-crowded blue end of the spectra. 

\subsection{Magnesium Abundances}

The high-resolution spectra analyzed in this study can be used to probe chemical evolution
using a number of different elements, and an analysis of Mg abundances, which is a product
of Supenovae Type II (SN II), is presented here as a comparison to the Fe abundances,
produced mainly in SN Ia.  Within the spectral interval covered here,
the Mg I line at 5711.090 \AA\ is well-defined and measurable.  This line
has a relatively accurate oscillator strength ($\sim$14\%) and was
used in the \citet{smith00} study of $\omega$ Cen.  
In the abundance analysis of Mg, the atmospheric parameters defined by the 
analysis of the Fe lines were used while the Mg abundance was varied to 
determine the best-fit profile to the Mg I line.

\section{Discussion}

\begin{figure}
\includegraphics[scale=.80]{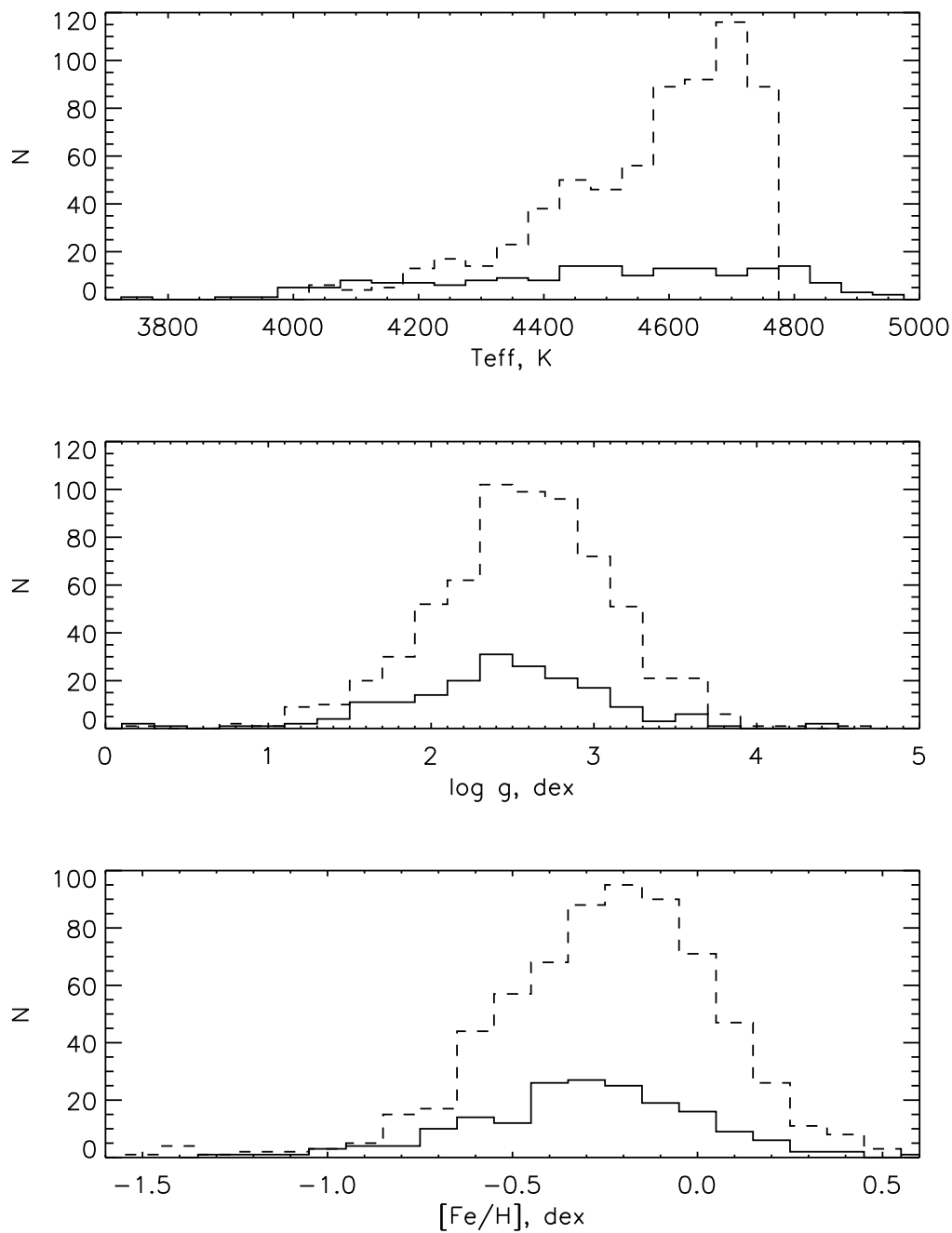}
%--%\epsscale{.70}
%\plotone{fig3.eps}
\caption{
Distributions of $T_{eff}$, $\log~g$, and [Fe/H]
for the GGSS (solid line) and Tycho (dashed line) samples. 
In terms of the average effective temperatures and surface gravities, the two
samples are similar ($<T_{eff}>$=4567 and 4503 with $<log g>$=2.56 and 2.42,
for Tycho and GGSS, respectively). The effective temperatures of the 
stars in the Tycho sample, however, are more localized as they are 
predominantly red clump stars. 
\label{fig3}
}
\end{figure}

Figure~\ref{fig3} shows the distribution of $T_{eff}$, $\log~g$, and [Fe/H]
%% -- , $\xi$, and $vsini$ 
for all sample stars. Since this study is comprised of two different samples,
the GGSS candidates are indicated as the solid line histogram, while the Tycho stars
are represented as the dashed line histogram. 
The top panel of Figure 5 shows clearly that the Tycho sample stars, which
were picked to be probable clump red giants, tend to cluster in
a narrower T$_{\rm eff}$ range, as would be expected for clump giants.  
The GGSS giants, on the other
hand, spread across the entire red giant branch, as evidenced
by the broad distribution in T$_{\rm eff}$. The middle panel shows the 
distribution of the samples in terms of their surface gravities. The average
$\log~g$ for the Tycho sample is 2.56 $\pm$ 0.55 and a similar average value is found
for the GGSS sample 2.42 $\pm$ 0.64. The bottom panel
illustrates the [Fe/H] distributions and shows the mean
subsolar metallicities of both Tycho and GGSS samples, with these
samples dominated by thick disk members. 
%The formal mean and r.m.s. scatter of the atmospheric parameters for the two samples is given in Table~\ref{tab5}.
%It should be noted that the only significant difference between the two samples is 
%the signal-to-noise ratio of the observations, which is systematically higher for the Tycho stars since 
%they are, on average, brighter than the GGSS stars. 

\begin{figure}
\includegraphics[scale=.80]{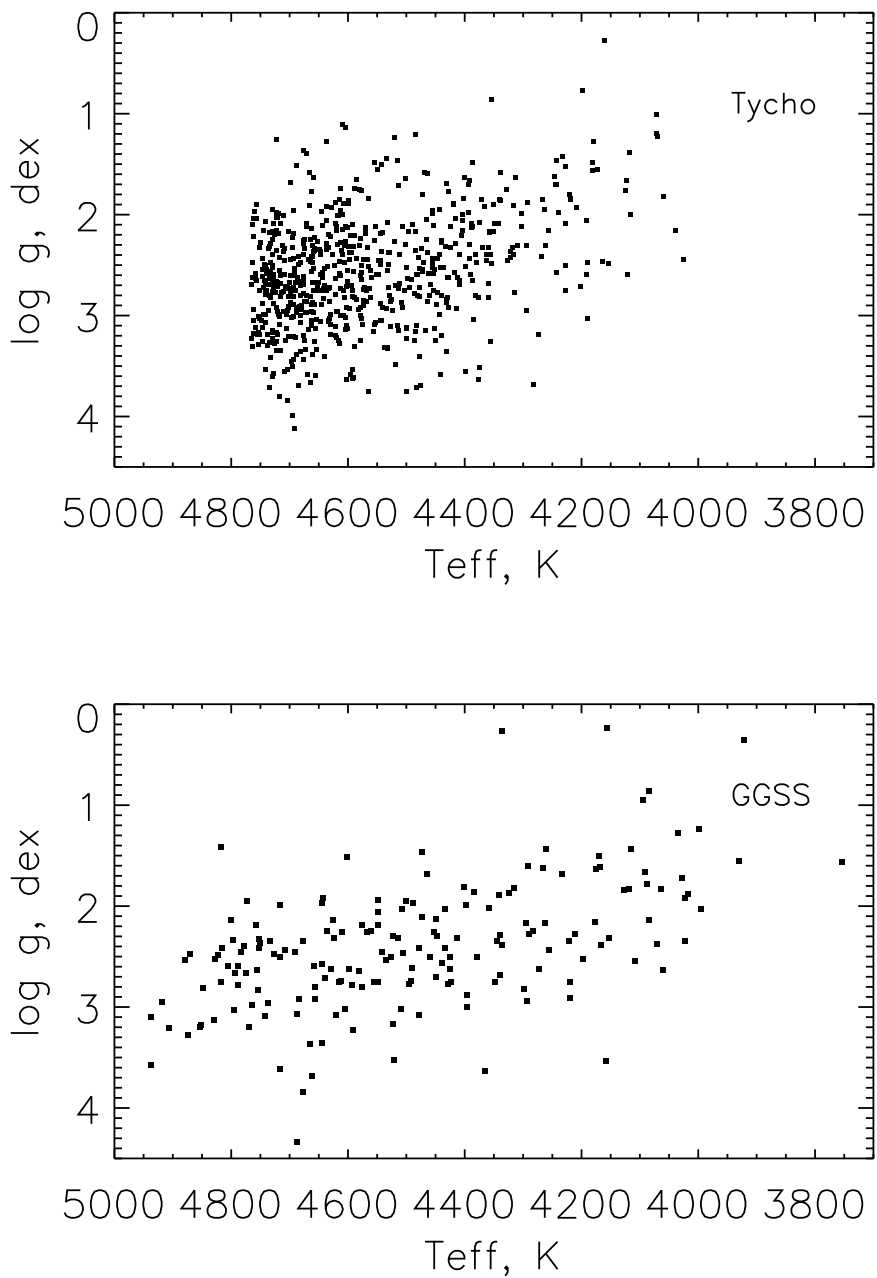}
%--%\epsscale{.70}
%\plotone{fig4.eps}
\caption{
The diagram $T_{eff}$ versus $\log~g$ for the Tycho (upper panel) and GGSS 
(lower panel) samples.  The distribution of the two different samples
in this diagram illustrates the clustering of the Tycho sample to a
more localized region in the T$_{\rm eff}$--log g plane when compared
to the GGSS sample.  The Tycho sample contains a sizable fraction of
red giant clump stars, while the GGSS sample consists of giants
distributed more uniformly across the first ascent of the red giant
branch.
\label{fig4}
}
\end{figure}

Figure~\ref{fig4} places the sample stars on a  diagram of $T_{eff}$ versus $\log~g$. 
Most of the stars are located along typical red giant branch and clump loci.
The lower panel in Figure~\ref{fig4} confirms one conclusion by \citet{b06} that the 
photometric GGSS survey \citep{patterson01} selected red giants very 
efficiently. The Tycho sample was selected with color cuts and has no stars
hotter than 4750 K in our sample of stars. The Tycho objects 
tend to concentrate in the location of the red giant clump in the T -- log~g 
diagram \citep[see][ for example]{b06}.

%% Table~5. Atmospheric parameters for the GGSS and Tycho samples.

%\begin{deluxetable}{ lcc }
%%\setcounter{table}{2}  
%\tablewidth{0pt}  
%\tablecaption{Mean atmospheric parameters for 
%the GGSS and Tycho samples.}
%
%\tablehead{  
%Sample & Tycho & GGSS \\
%} 
%\startdata
%N stars       &         658         &           183         \\   
%$T_{eff}$, K  &  4567  $\pm$  158   &    4502  $\pm$  278   \\        
%$\log~g$, dex &  2.56  $\pm$  0.55  &    2.42  $\pm$  0.64  \\
%$[Fe/H]$, dex & -0.24  $\pm$  0.30  &   -0.30  $\pm$  0.31  \\
%$\xi$, km/s   &  1.30  $\pm$  0.28  &    1.31  $\pm$  0.28  \\
%$vsini$, km.s &  1.67  $\pm$  1.26  &    2.14  $\pm$  1.46  \\
%S/N           &   43   $\pm$  12    &      34  $\pm$  11    \\
%\enddata
%\tablecomments{Number of spectra, effective temperature, surface gravity, 
%metallicity, microturbulence velocity, stellar rotation, and signal-to-noise
%for Tycho and GGSS samples.}
%\label{tab5}
%\end{deluxetable}

In Figure 7 we plot the values of
[Mg/Fe] versus [Fe/H] for the high-S/N sample of the brighter giants
and note that the well-established increase of [Mg/Fe] with decreasing
[Fe/H] is recovered in this automated analysis.  As an illustration
of the agreement between this study and previous work, the abundances
from Mishenina et al. (2006), who analyzed nearby bright clump giants,
are also included in Figure 7.  The general agreement demonstrates that
the automated analysis techniques developed here produce reasonable
abundance results, at least from spectra of relatively high S/N.

Of more interest to continuing abundance studies based on these spectra
are the Mg abundances derived from the entire sample, which generally
consists of spectra with lower values of S/N.  The abundance values of
[Mg/Fe] versus [Fe/H] for all of the GGSS and Tycho giants are shown in
Figure 8; here again, the well-established trend of increasing [Mg/Fe]
with decreasing [Fe/H] is clearly evident.  As a comparison study
which is based on high-S/N spectra and employs a classical analysis, we
include in Figure 8 the abundances from Reddy et al. (2003; 2006), which
sampled thick and thin disk stars.  This is an appropriate comparison
since the majority of giants in this study are likely to be members of
the thick disk.  The summary point from Figures 7 and 8 is that this
set of spectra can be used to probe the chemical evolution of a number
of elements.

\section{Conclusions}

We have developed a pipeline to estimate the stellar 
atmospheric parameters, metallicities and magnesium abundances for a large set of
spectra of candidates to the SIM Planet Quest reference frame stars 
observed with an echelle spectrograph in 2004-2006.  Most of the stars photometrically preselected as
red giants turn out to be bona fide red giants with slightly subsolar metallicities,
on average. 
The Tycho sample stars were observed with higher S/N, on average, and are
more concentrated to the loci of red giant clump stars than the GGSS sample. 
The latter one consists of giant stars with $T_{eff}$ uniformly distributed
between 4000 K to 5000 K.
The iron abundances of the GGSS and Tycho samples are similar, with
average values of [Fe/H] = -0.30 and -0.24, respectively. This average
is more metal-poor than the value for the thin disk. 

The values of [Mg/Fe] versus [Fe/H] in the GGSS and Tycho program stars 
derived from the automated analysis developed here when compared to results
from the classical analysis in \citet{reddy03,reddy06} is very good, demonstrating
that the automated analysis produces reliable abundances for Fe and Mg.
The next step in the automated analysis of this spectral dataset, which
will be the topic of a future paper, will be abundance determinations of
additional elements, with the particular elements chosen in order to
study chemical evolution in this sample of predominantly thick and
thin disk giants.

\begin{acknowledgments}

We are grateful to our collaborators who helped with observations: 
William Cohran, Mike Endl, Nairn Baliber (all from UT Austin), and 
Nataly Petrova (McDonald Observatory). 
The GGSS follow-up observations presented here were
supported by the JPL and NASA via grant 99-04-OSS-058.
We are grateful to the staff of McDonald observatory for their help and
support in photometric and high resolution spectroscopic observations.
This publication makes use of data products from the Two Micron All Sky
Survey, which is a joint project of the University of Massachusetts and the
IPAC/Caltech, funded by the NASA and NSF.

\end{acknowledgments}

%%\clearpage

%%% Fig.1

%%% Fig.2

%%% Fig.3

%%% Fig.4

%%% Fig.5

%%% Fig.6

%%% Fig.7
\begin{figure}
\includegraphics[scale=.80]{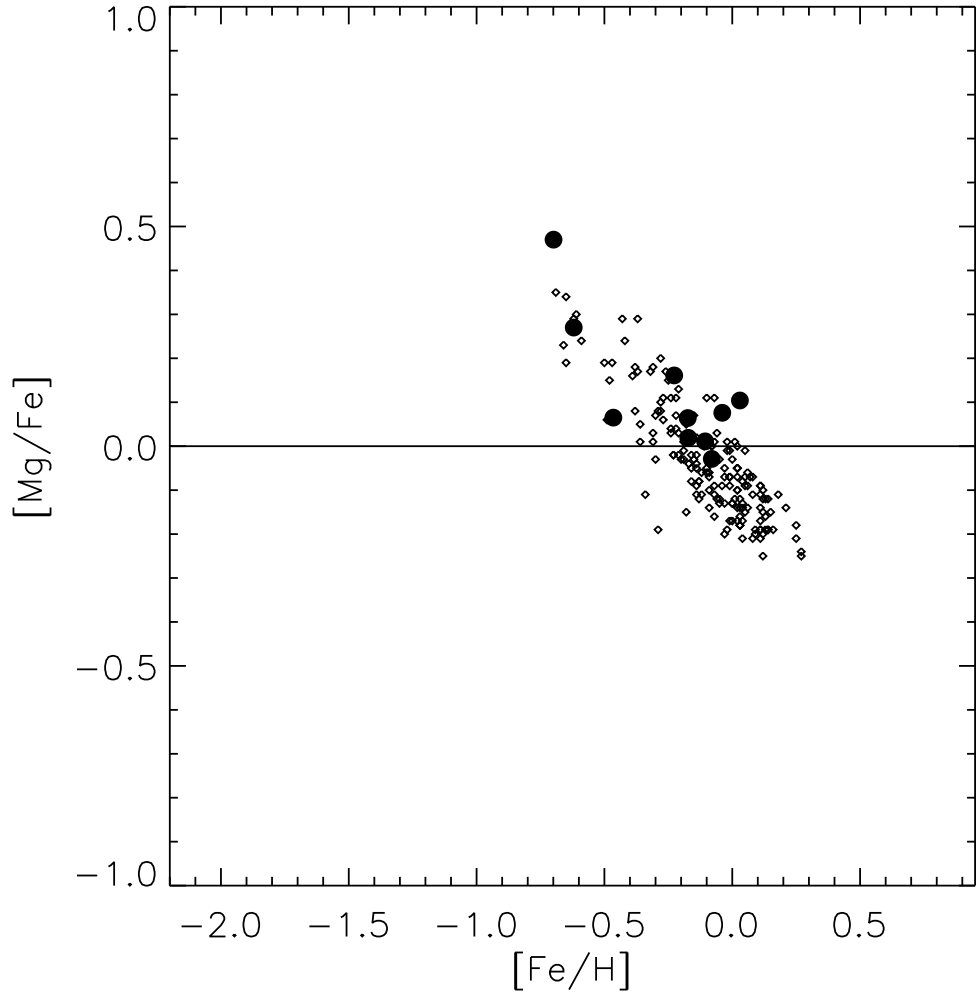}
%--%\epsscale{.70} 
%\plotone{fig4.eps}
\caption{
The ratio of Magnesium/Iron abundances versus iron abundances in the 
control sample of 10 bright stars (filled circles), 
with the increase in [Mg/Fe] with decreasing [Fe/H]
clearly evident.  Included in this figure are abundance results from
Mishenina et al. (2006) shown with diamonds and the agreement 
between the two studies is excellent.
\label{Mg1plot}
}
\end{figure}

%%% Fig.8
\begin{figure}
\includegraphics[scale=.80]{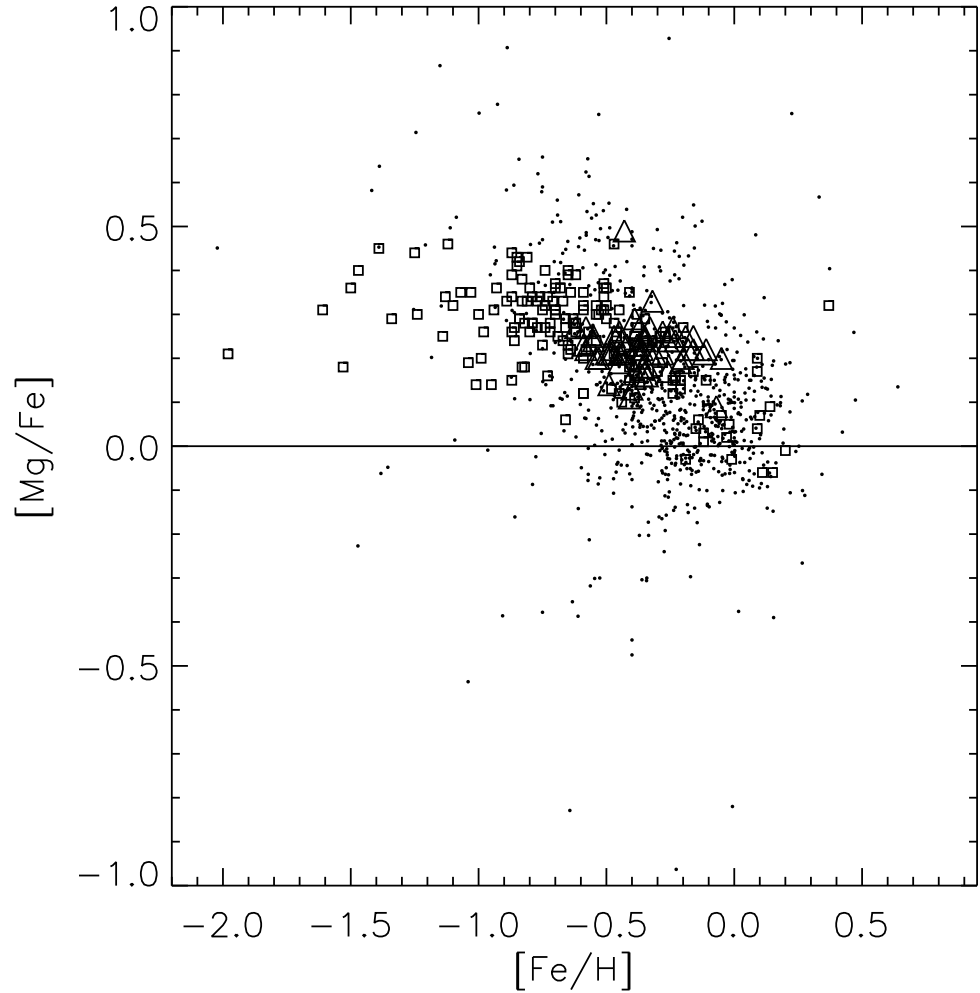}
%--%\epsscale{.70}
%\plotone{fig4.eps}
\caption{
The values of [Mg/Fe] versus [Fe/H] in the GGSS and Tycho program stars (dots).
The squares and triangles show the abundance results from 
\citet{reddy03,reddy06}, respectively.  The agreement between the abundances
derived from the automated analysis developed here when compared to the
classical analysis from \citet{reddy03,reddy06} is very good, demonstrating
that the automated analysis produces reliable abundances for Fe and Mg.
\label{Mg2plot}
}
\end{figure}

\end{document}